\documentclass[prx,twocolumn,floatfix,superscriptaddress,amsmath,caps,longbibliography]{revtex4-2}
\pdfoutput=1
\usepackage{dcolumn,graphicx,color,booktabs,microtype,afterpage} 
\makeatletter\renewcommand{\fnum@figure}[1]{\figurename~\thefigure.}\makeatother
\makeatletter\renewcommand{\fnum@table}[1]{\tablename~\thetable.}\makeatother
\usepackage[colorlinks,plainpages=false,linkcolor=blue,urlcolor=blue,citecolor=blue,pdfpagemode=UseNone,pdfstartview=FitBH]{hyperref}
\usepackage{bm} 

\newcommand{\TFO}{TbFeO$_3$}

\begin{document}
	
\title{Spin dynamics and exchange interaction in orthoferrite TbFeO$_{3}$ with non-Kramers rare-earth ion}

\author{S.~A.~Skorobogatov}
\affiliation{Kirensky Institute of Physics, Federal Research Center KSC SB RAS, Krasnoyarsk, 660036 Russia}
\affiliation{Siberian Federal University, Krasnoyarsk 660041, Russia}
\author{K.~A.~Shaykhutdinov}
\thanks{Corresponding author: smp@iph.krasn.ru}
\affiliation{Kirensky Institute of Physics, Federal Research Center KSC SB RAS, Krasnoyarsk, 660036 Russia}
\affiliation{Siberian Federal University, Krasnoyarsk 660041, Russia}
\author{D.~A.~Balaev}
\affiliation{Kirensky Institute of Physics, Federal Research Center KSC SB RAS, Krasnoyarsk, 660036 Russia}
\affiliation{Siberian Federal University, Krasnoyarsk 660041, Russia}
\author{M.~S.~Pavlovskii}
\affiliation{Kirensky Institute of Physics, Federal Research Center KSC SB RAS, Krasnoyarsk, 660036 Russia}
\affiliation{Siberian Federal University, Krasnoyarsk 660041, Russia}
\author{A.~A.~Krasikov}
\affiliation{Kirensky Institute of Physics, Federal Research Center KSC SB RAS, Krasnoyarsk, 660036 Russia}
\author{K.~Yu.~Terentjev}
\affiliation{Kirensky Institute of Physics, Federal Research Center KSC SB RAS, Krasnoyarsk, 660036 Russia}

\begin{abstract}

    The low-temperature spin dynamics of the orthorhombic TbFeO$_3$ perovskite has been studied. It has been found that the inelastic neutron scattering (INS) spectrum contains two modes corresponding to different sublattices in the compound. The iron subsystem orders antiferromagnetically at T$_\mathrm{N} =$ 632~K and exhibits the high-energy magnon dispersion. Magnetic dynamics of this subsystem has been described using the linear spin wave theory and our solution yields sizable anisotropy between in-plane and out-of-plane exchange interactions. This approach was previously used to describe the magnon dispersion in the TmFeO$_3$ compound. Three non-dispersive crystal electric field levels corresponding to Tb$^{3+}$ ions have been established in the region below 40 meV at about 17, 26, and 35~meV. Study of diffuse scattering at different temperatures has elucidated the behavior of the magnetic correlation length. The behavior of the Tb$^{3+}$ ion subsystem has been numerically described in the framework of the point charge model. The numerical data agree satisfactorily with the experiment and with the general concept of the single-ion approximation applied to the rare-earth subsystem of orthorhombic perovskites.
    
\end{abstract}

\maketitle{}

\section{Introduction}

The magnetic oxide materials containing transition and rare-earth ions exhibit a diversity of intriguing effects related to the complex interaction between two magnetic subsystems. Among such materials are manganites with the general formulas RMnO$_3$ ~\cite{1, 2} and RMn$_2$O$_5$ ~\cite{3, 4} and huntites RFe$_3$(BO$_3$)$_4$ ~\cite{5, 6, 7, 8}. A separate class of the oxide compounds are orthoferrites with the general formula RFeO$_3$ ~\cite{9}, which have been explored for several decades ~\cite{10, 11, 12}. The crystal structure of the RFeO$_3$ compound is described by space group \textit{Pbnm}. The recent renewed interest in these compounds is due to the discovery of several curious phenomena, including multiferroism below the temperature of ordering of the rare-earth subsystem ~\cite{13}, laser-induced ultrafast magnetization switching of domain walls ~\cite{14, 15, 16}, and the formation of a soliton lattice in the TbFeO$_3$ compound ~\cite{17}.

The family of rare-earth orthoferrites is attractive in terms of magnetic phenomena. The unique magnetic properties of these materials are caused by complex interactions between the magnetic moments of 3d and 4f electrons. As was shown in ~\cite{9}, the RFeO$_3$ compounds have extraordinarily high Néel temperatures ($T_{\rm N}$ $\approx$ 600–700 K), below which the moments of iron are ordered antiferromagnetically with a weak canting of the sublattices resulting in weak ferromagnetism. As the temperature decreases, the Fe‒R coupling becomes more significant and induces spin-reorientation transitions at $T$ $\ll$ $T_{\rm N}$. The temperature $T_{\rm SR}$ of the spin-reorientation transition depends strongly on a rare-earth ion. In particular, in the HoFeO$_3$ compound, this temperature is $T_{\rm SR}$ $\approx$ 50‒60 K; for TmFeO$_3$, it is $T_{\rm SR} \approx\ 80‒90$~K~\cite{18}; while in the case of Tb as a rare-earth element, this temperature is much lower: $T_{\rm SR}$ $\approx$ 3‒10 K ~\cite{17, 19}. At high temperatures, the subsystem of rare-earth ions with a relatively weak R‒R coupling is paramagnetic or weakly polarized by the molecular field of ordered Fe ions. The rare-earth magnetic sublattice is ordered at different temperatures, depending on the type of an R ion and external conditions. The complex magnetic properties of this system are determined by the diversity of exchange interactions. Here, along with the Heisenberg-type Fe–Fe, Fe–R, and R–R exchange interactions, an important role is played by the Dzyaloshinskii–Moriya (DMI) interaction ~\cite{20}, which induces a weak ferromagnetic moment.

Previously, several orthoferrites were investigated~\cite{21} and isostructural perovskite compounds with different rare-earth ions using inelastic neutron scattering as a basic research tool for studying the energy spectra of excitations in the magnetic subsystems. In the YbFeO$_3$ compound~\cite{22}, the authors found quasi-one-dimensional Yb$^{3+}$ chains and ``shadow modes'' in the low-temperature spin dynamics of the rare-earth subsystem. Study of the related YbAlO$_3$ compound ~\cite{23, 24} disclosed unique magnetic quantum effects in the rare-earth subsystem at ultralow temperatures. It should be noted that Yb$^{3+}$ is a Kramers ion, the ground state of which is split into doublets; therefore, the pseudospin 1/2 model is applicable. Upon splitting of the ground state in non-Kramers ions, the latter can create various combinations of singlet, doublet, pseudo-doublet, and other combination of energy levels. A more complex picture of the energy states significantly complicates the description of the behavior of the rare-earth subsystem and the 3d‒4f interaction arising between the sublattices, but the intriguing states and effects that can be observed in the compounds containing such ions broaden the scope for the development of physics of oxide materials.
Another compound studied was TmFeO$_3$ ~\cite{25}, in which we observed the effect similar to the ``shadow mode'' of the rare-earth subsystem and the spin dynamics of the second singlet level. This subsystem was described in the approximation of an ion in an external field. We determined several parameters that can be used in the magnetic Hamiltonian, although a variety of magnetic states of the Tm$^{3+}$ ion make this problem challenging both numerically and analytically.
In this study, to solve the problem of non-Kramers rare-earth ions in orthoferrites, we examined the TbFeO$_3$ compound. This material has two ordering temperatures: $T_{\rm N}^{\rm Fe}$ $\approx 650$~K and $T_{\rm N}^{\rm Tb}$ $<$ 10 K \cite{17, 19}. Below 8 K, it undergoes a spin-reorientation transition. This compound was investigated using the approaches applied previously to TmFeO$_3$ to better understand physics of the interactions in rare-earth orthoferrites.

\section{EXPERIMENTAL DETAILS} 

\begin{figure}[tb!]
	
	\includegraphics[scale=0.132]{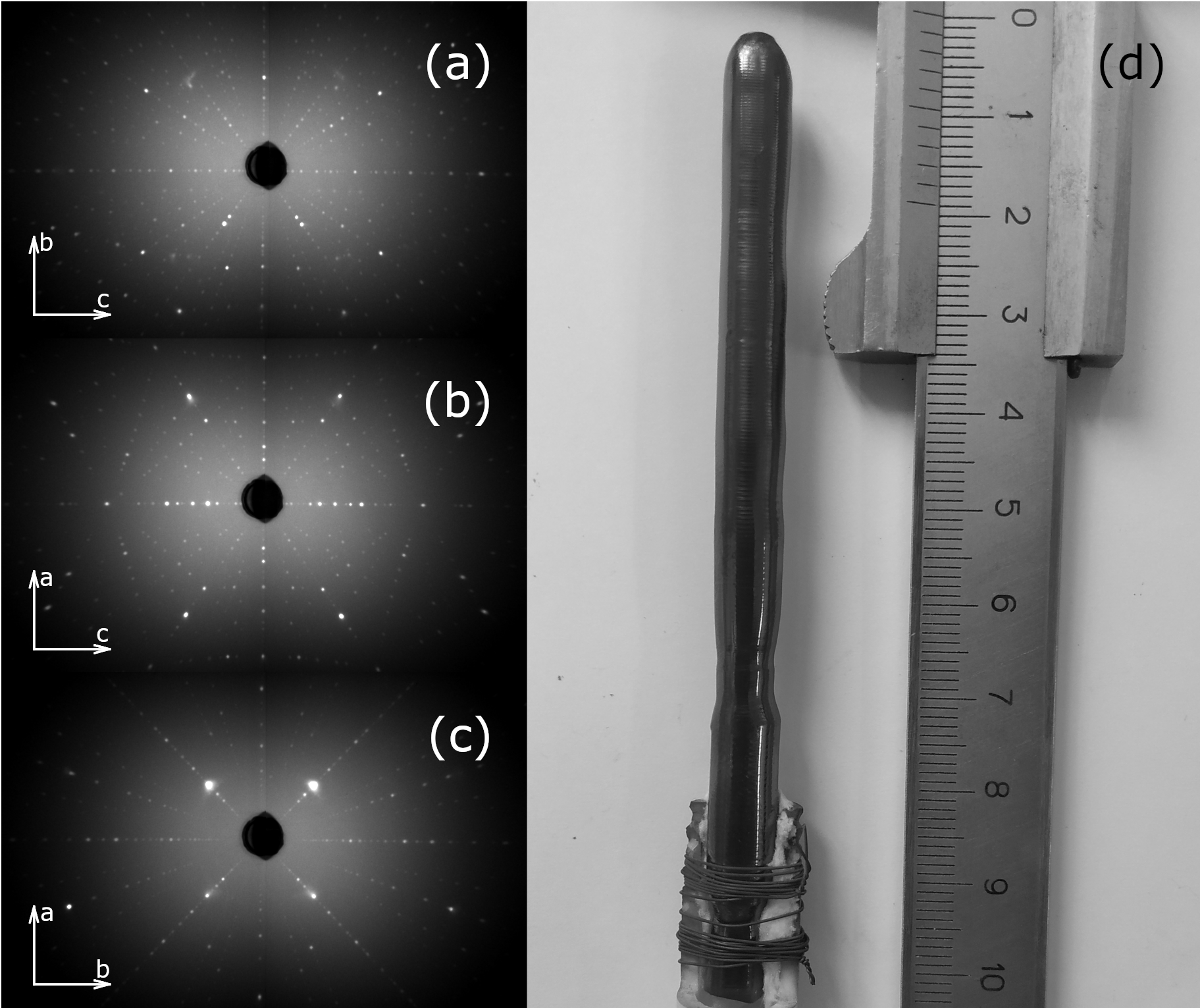}
	\caption{ (a, b, c) Lauegrams of the TbFeO$_3$ single crystal along different crystallographic directions corresponding to sp. gr. Pbnm with lattice parameters of $a = 5.326$~\AA, $b = 5.602$~\AA, and $c = 7.635$~\AA. (d) Photograph of the single crystal obtained by optical floating zone melting.}\label{Laue} 

\end{figure}

The TbFeO$_3$ single crystal was grown in a Crystal Systems FZ-T-4000-H optical floating zone furnace (Japan) at a growth rate of 3 mm/h and a relative rod rotation at 30 rpm in air under ambient pressure. A sample 5×5×5 mm$^3$ in size was cut from the grown rod along its crystallographic axes. The quality of the single crystal was estimated by the Laue method. The Lauegrams and photograph of the grown crystal are presented in Fig ~\ref{Laue}.

The magnetic measurements were performed on a Quantum Design PPMS-6000 Physical Property Measurement System. The temperature and field dependences of the magnetization were measured along different crystallographic directions in the single-crystal sample.

\begin{figure}[tb!]
	
	\includegraphics[width=1.0\linewidth]{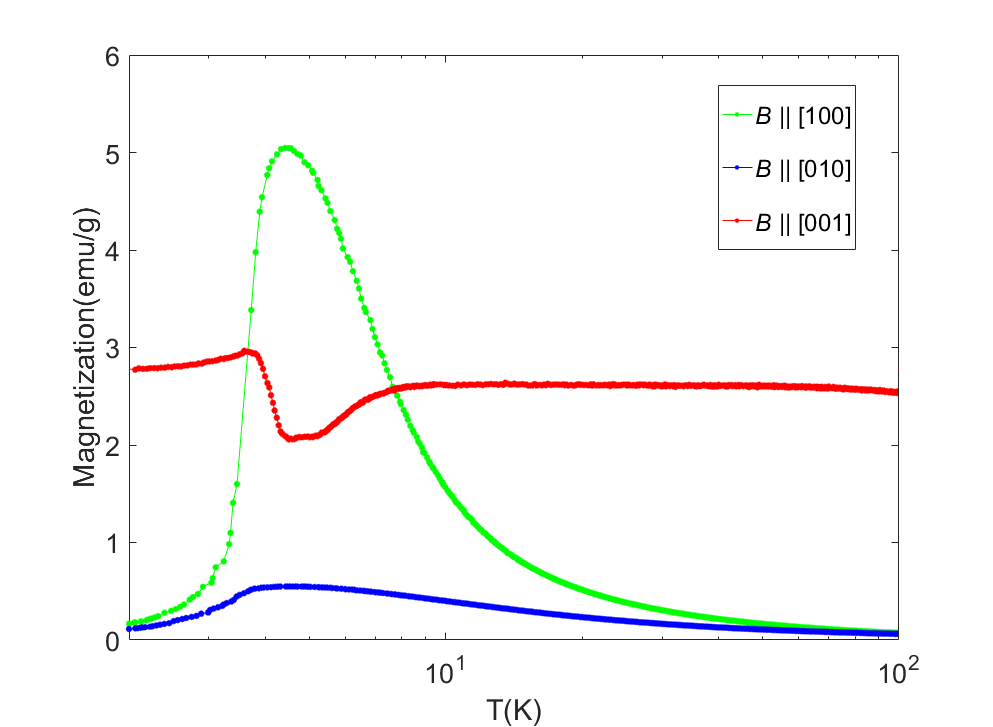}
	\caption{ Temperature dependence of the magnetization in logarithmic scale. The measurements were performed at temperatures from 2 to 100 K after pre-cooling to 2 K without external magnetic field. At 2 K, the hysteresis was measured in fields of $~\pm$5 T; after that, the field was decreased from 5 to 0.01 T and the temperature dependence of the magnetization was measured. Of particular interest is the temperature range from 3.5 to 4.5 K corresponding to the spin-reorientation transition.}\label{M(T)} 
    \vspace{-10pt}
\end{figure}

The behavior of the $M(T)$ dependences presented in Fig. ~\ref{M(T)} discloses a spin-reorientation transition in the temperature range of 3.5–4.5 K. The transition of the terbium subsystem is not observed under the experimental conditions (Fig. ~\ref{M(T)}). In ~\cite{17}, this transition was detected at 7–8 K. It is noteworthy that the inelastic neutron scattering experiment used as a source of the main conclusions drawn in this study was carried out at $T = 2$~K, i.e., below the temperature region of the observed spin-reorientation transition.
To comprehensively analyze excitations of both the iron and terbium subsystems in the TbFeO$_3$ single crystal, the inelastic neutron scattering study was carried out over a wide range of the reciprocal space and energy transfer. The experiments were conducted using two time-of-flight mass spectrometers: an angular-range chopper spectrometer (ARCS) ~\cite{27} and a cold neutron chopper spectrometer (CNCS) ~\cite{28, 29} at the Oak Ridge National Laboratory’s Spallation Neutron Source. The data were obtained on a single crystal with a mass of 3 g oriented in the scattering plane $(0KL)$. The incident neutron energies were $E_{\rm i}$ = 100 and 25 meV for the ARCS and $E_{\rm i}$ = 12 and 3.3 meV for the CNCS.
The experimental data were processed and preliminary analyzed using the Dave ~\cite{30}, Horace ~\cite{31}, and MantidPlot ~\cite{32} software. The excitation spectra and neutron scattering cross sections of the spin Hamiltonian were calculated using the linear spin wave theory (LSWT) implemented in the SpinW software package ~\cite{33}. The crystal electric field (CEF) was calculated using the McPhase software ~\cite{34}.

\section{RESULTS AND ANALYSIS}

\subsection{Iron subsystem} 

In this section, we analyze the data obtained on the high-energy time-of-flight ARCS at incident neutron energies of 150 and 50 meV. The cuts made along the highly symmetric directions of the reciprocal space reveal the magnon dispersion, which attains its maximum energy of 60 meV and can be considered with confidence to be the dispersion of the iron subsystem ~\cite{22, 25, 35, 36}. The low-energy region was investigated at neutron energies of up to 25 meV, which made it possible to examine the antiferromagnetic gap of the iron subsystem and determine the general characteristics of the rare-earth subsystem.

\begin{figure}[t]
	
	\centering\includegraphics[scale=0.3]{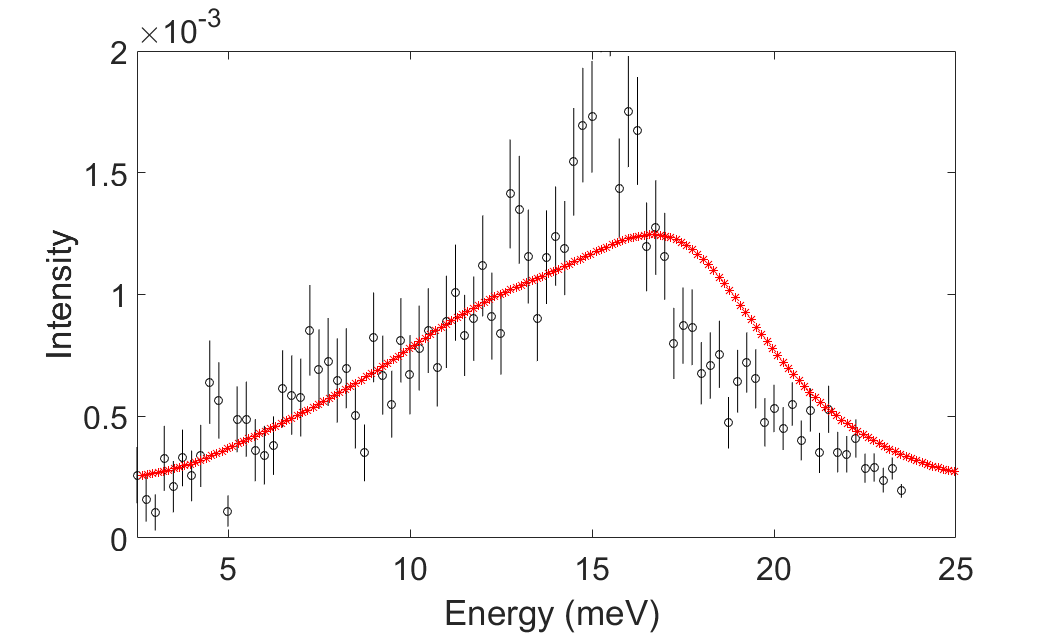}
	\caption{ Energy slice made near $\Gamma$ point [0,1,3]. Experimental data were received summation on high-symmetry directions, intergration limit in $\pm$~0.1 r.l.u. Experimental data(black points with errorbars) and results what we were calculated with our model(red points). On figure we highlighted some area, at the first, 4-5 meV located here experimental point  indicated about end of AFM gap and start magnon dispersion, second, near 7~meV we think it is splitting of magnon mod caused by DM interaction and thirdly large area 14-17 meV here located CEF levels which we discuss below in the main text.}\label{Fe magnon gap} 

\end{figure}

Using the experimental data, we determined positions of the curves at 72 points of the reciprocal space along the three highly symmetric directions in order to determine the spin Hamiltonian. The main challenge was to find an error of the selected points. The attempts to describe the selected points by the Gaussian function did not yield objective results because of the low density of points in the investigated ranges and a large error. Therefore, we chose another model for determining an error of the points. The algorithm used was iterative and included the operation of comparison of the intensities of the initially proposed point and its neighbors in the chosen $Q$ or $E$ direction, depending on the cut. Next, if the ratio between intensity $I_1$ of the selected point and neighbor’s intensity $I_2$ was less than a chosen value, the process was repeated for the next neighbors with the accumulation of the error factor n; otherwise, the solution was considered to be found and an error equal to 1/2 of the step in the cut direction $*n$ was assigned to the value. Using the described procedure, we found one of the parameters (the iron antiferromagnetic gap) with good accuracy and determined the anisotropy constant $K_c$ ~\cite{37} (Fig. ~\ref{Fe magnon gap}).

The behavior of the iron subsystem was described using the SpinW software package ~\cite{33} and the classical spin Hamiltonian 

\begin{equation}\label{Hamiltonian}	
	H=\sum_{i,j}J_{i,j}S_iS_j+K_c\sum_{i}S_{c_i}^2	
\end{equation}
in which two terms were taken into account: the first term is the Heisenberg exchange interaction between neighboring ions and the second term is the effective anisotropy. The DMI causes splitting of the magnon modes in the vicinity of the Gamma point ($\Gamma$)~\cite{21}; however, the presence of the CEF excitation of the terbium subsystem around 17 meV makes the signal noisy in this region, so the effect of the DMI on the spin dynamics cannot be reliably established. The first summation is made over different sets of neighbors of Fe atoms. In Eq. (\ref{Hamiltonian}), $K_c$ is the effective anisotropy responsible for the stabilization of the $\Gamma_2$ phase. The exchange interaction model used by us is similar to that chosen in ~\cite{25}; the in-plane and out-of-plane exchanges are separated. By fitting the model to experimental points, we obtained exchange values of $J_ab$ = 4.74, $J_c$ = 4.96, $J_{ab2}$ = 0.02, and $J_c2$ = 0.2, an anisotropy value of $K_c$ = ‒0.20, and deviations of $R_w$ = 2.51.

\begin{figure}[t]\label{Fe magnon mods}
	
	\centering\includegraphics[scale=0.147]{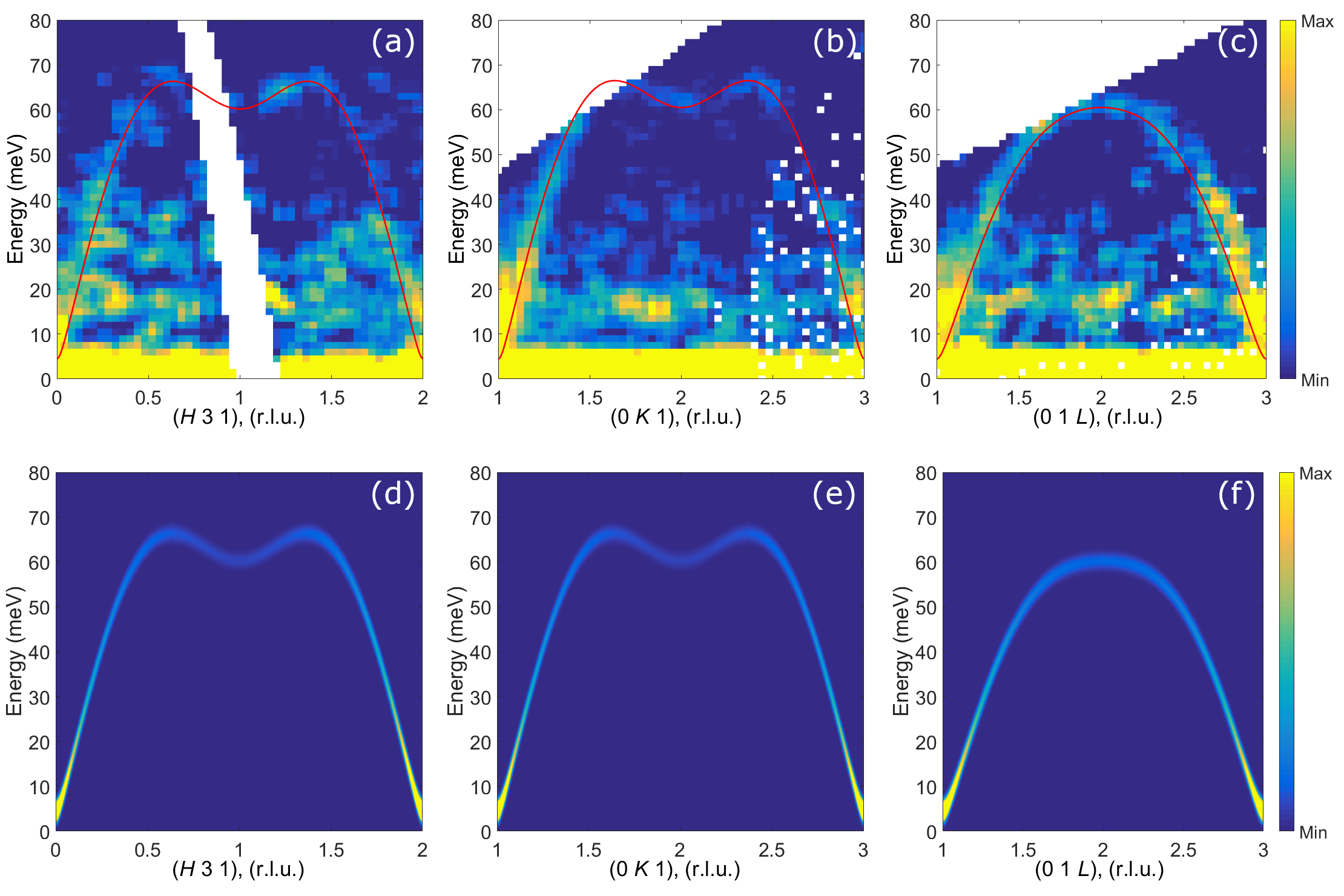}
	\caption{ Cuts along the three characteristic directions of the reciprocal space. (a‒c) Experimental inelastic neutron scattering spectra. The measurements were performed at a temperature of 2 K. The data were obtained by summation of the symmetric directions. The integration limit is $\pm$0.1 r.l.u. A gaussian smoothing was applied to the data. The calculated dispersion curves (red) are superimposed. (d‒f) Model calculation of the inelastic scattering spectra using the selected model.}
\end{figure}

In addition, this approach allowed us to get rid of the non-representativeness of errors, but required the more thorough selection of fitting results in the physical model used. This was expressed in the variance of the solutions found; specifically, two stable solutions appeared (see Table I). The analysis of our data can be considered satisfactory, since we managed to obtain the exchange values for the Fe$^{3+}$ subsystem and thereby characterize the magnon dispersion in all the three directions of the reciprocal space (Fig. ~\ref{Fe magnon mods}). Good agreement between the experimental and calculated dispersions indicated the applicability of the LSWT-based model. In addition, the results on the iron subsystem obtained in this study are consistent with the data reported in ~\cite{25}.

\

\tablename{ I. Stable exchange solutions for the iron subsystem}

\begin{center}
\begin{tabular}{lcccccc}

	\hline\hline
	
	  & $J_{ab}$ & $J_c$ & $J_{2ab}$ & $J_{2c}$ & $K_c$ & $R_w$ \\
	
	\hline
	
	First & 4.74 & 4.96 & 0.02 & 0.2 & -0.2 & 2.51 \\
	
	Second & 4.73 & 4.30 & 0.02 & 0.08 & -0.2 & 2.50 \\
	
	\hline
    
\end{tabular}
\end{center}

\

\subsection{Low-energy excitations}

\begin{figure}[h!]
	
	\centering\includegraphics[width=1\linewidth]{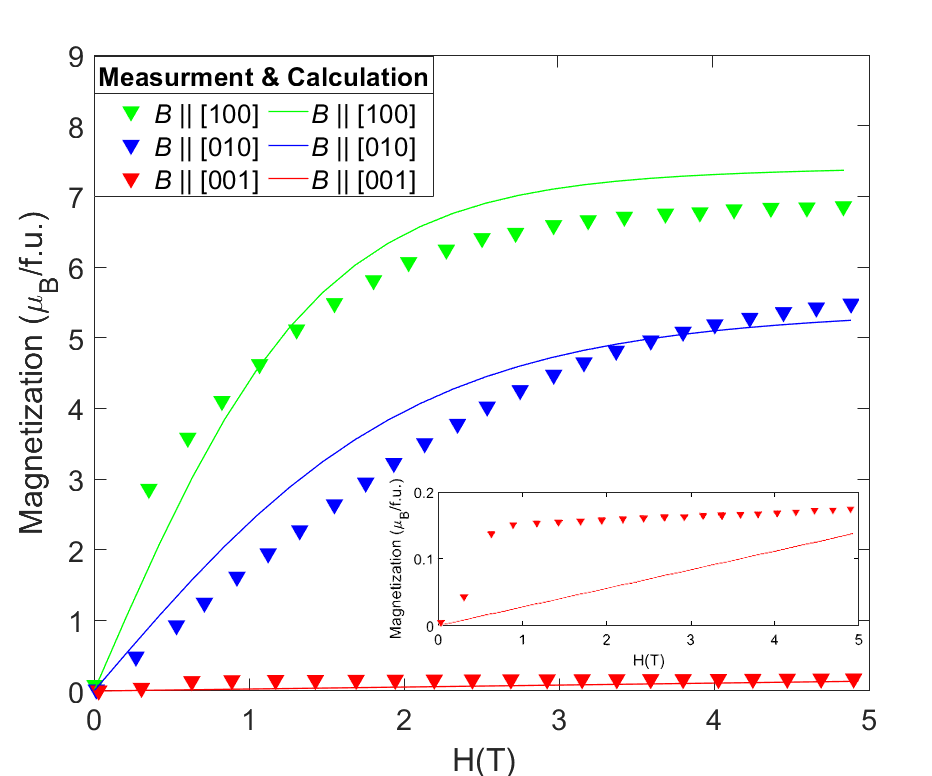}
	
	\caption{ Magnetization vs. magnetic field at 7 K. The [100], [010], and [001] crystallographic directions are colored in green, blue, and red, respectively. The solid lines correspond to the experiment and the triangles to the calculation.}\label{Magnetization} 
	
\end{figure}

The rare-earth Tb$_{3+}$ ion in the TbFeO$_3$ compound is a 4f element and its electrostatic interaction with surrounding ions can be described using Stevens operators. Terbium ions in \TFO\ have monoclinic local symmetry $C_s$ which yields 15 non-zero operators that need to be taken into account in the Hamiltonian ~\cite{36}.

\begin{equation}\label{CEF Hamiltonian}	
	H_{cef}=\sum_{l,m}B_l^mO_l^m		
\end{equation}
These parameters of the CEF can be determined using the point charge model (PCM). The experience in using this model was demonstrated in ~\cite{21, 32}. In the PCM, atoms localized in the selected unit volume are considered to be point charges in the crystallographic sites; this environment affects electrically the selected ion (in our case, Tb$^{3+}$) and, in this approach, any special interactions, spin parameters, etc. are ignored. We investigated neighboring ions lying within a sphere with a radius of $r$ = 4 \AA\ and calculated the parameters $B^m_l$ (a set of the calculated $B^m_l$ values is presented in Appendix, Table~A1) in the McPhase software package. Using the set of parameters, we simulated the CEF splitting, transition intensities, and magnetic anisotropy. The magnetization calculated using the obtained operators is shown in Fig. ~\ref{Magnetization}. It can be seen that the experimental and calculated data agree satisfactorily for all the three directions. The obtained experimental magnetizations are consistent with the data reported in ~\cite{25} and the comparison of the calculated and experimental curves confirms the applicability of the PCM in describing the behavior of the magnetization of rare-earth orthoferrites reported previously in ~\cite{22, 25}.

\begin{figure}[h!]
	
	\centering\includegraphics[width=1\linewidth]{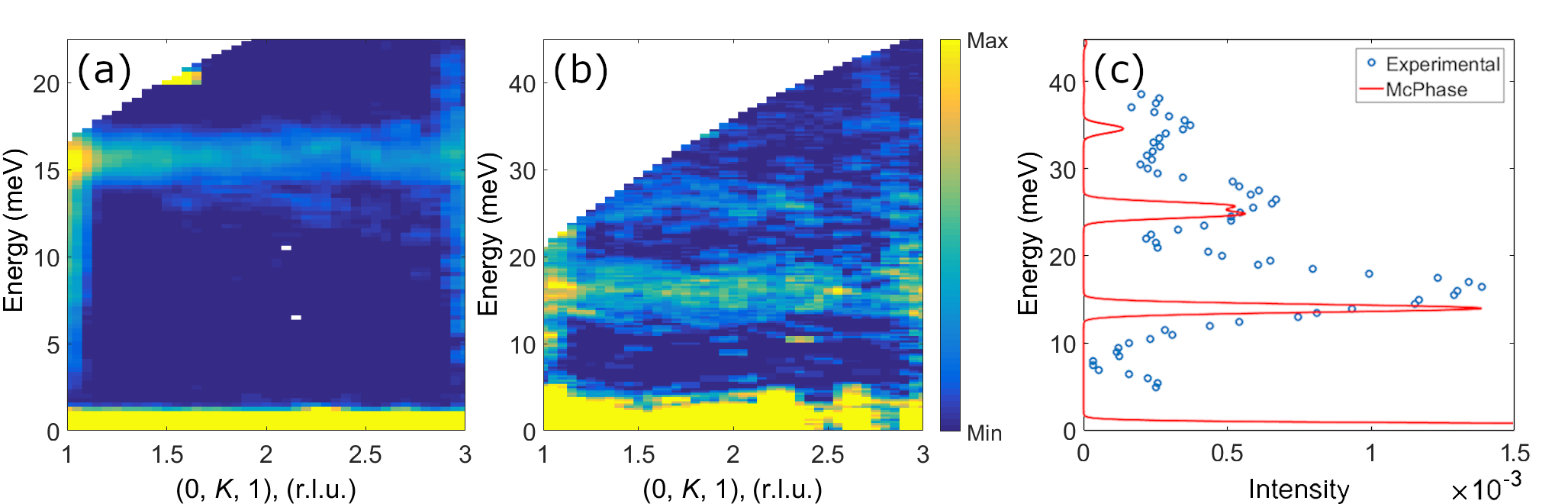}
	
	\caption{ (a, b) Cuts along the highly symmetric directions. The measurements were performed at 2 K. The data were obtained by summation of the symmetric directions. The integration limit is ±0.1 r.l.u. The spread operation was also applied. The crystal field level at 16‒17 meV is clearly seen in (a). In (b), there are two crystal field levels at 16‒17 and 26 meV; around 34 meV there is a hint of another crystal field level, but it cannot be established with confidence. (c) Comparison of the integral crystal field values with the calculation using the single-ion model in the McPhase software. The experimental data and the spin dynamics calculated using the PCM are in good agreement. The integration limit is ±0.25 r.l.u.}\label{CIF} 
	
\end{figure}

Based on the parameters calculated using the PCM, the transition energies for the Tb$^{3+}$ ion were also simulated. The ground state of this ion is degenerate and consists of 13 states, 6 of which are quasi-doublets and one is a singlet (a set of parameters is presented in Appendix). The first calculated quasi-doublet has an energy of E$_{0,1}$ $\sim$ 0 meV; this level belongs to the elastic neutron spectrum. The second CEF level is well-determined from the experimental data and located at ~17 meV; in the calculation, this level is represented by a quasi-doublet with E$_{2,3}$ $\sim$ 14 meV. Fig. ~\ref{CIF}(a), ~\ref{CIF}(b) show these energy levels together with the overlying levels corresponding to the quasi-doublets calculated by us using the PCM. Fig. ~\ref{CIF}(c) compares the calculated neutron dispersion intensities and the results of summation of the experimental data in the vicinity of the region [0, 2, 1]. The integration limit is ±0.25 r.l.u.

\begin{figure}[h!]
	
	\begin{minipage}[h]{0.485\linewidth}
	    \centering\includegraphics[width=1\linewidth]{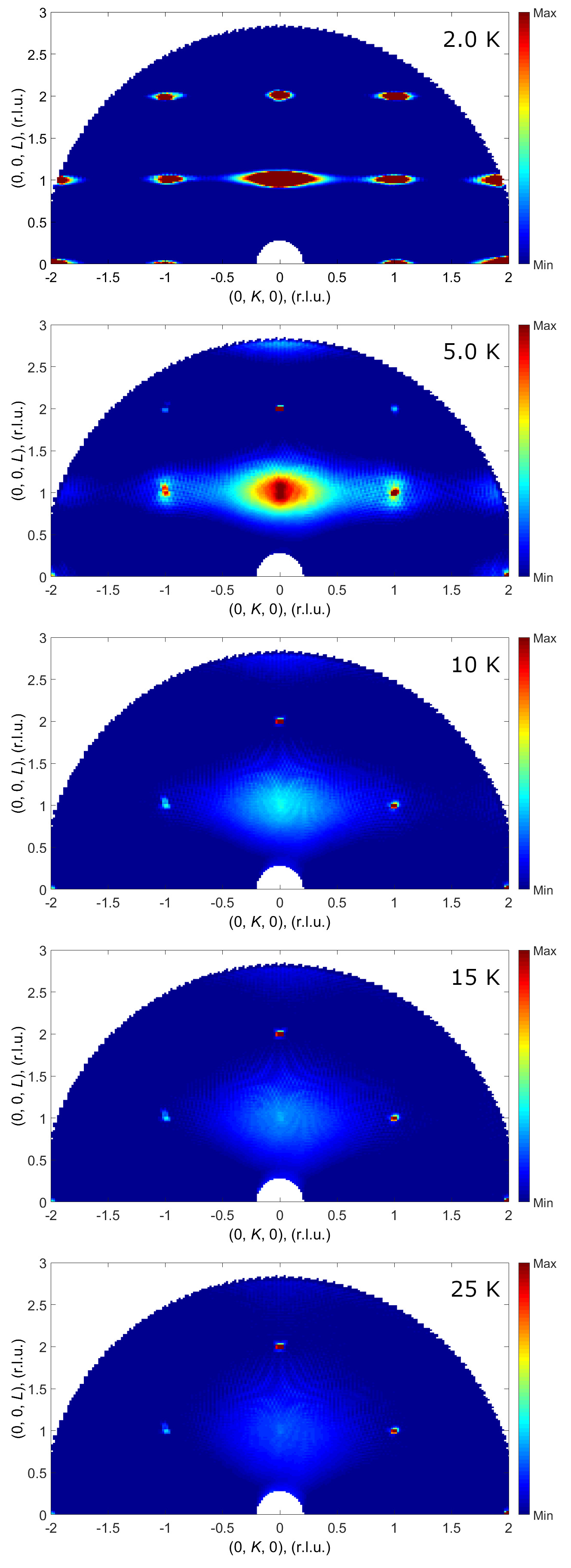} a\\
	\end{minipage}
	\hfill
	\begin{minipage}[h]{0.491\linewidth}
		\centering\includegraphics[width=1\linewidth]{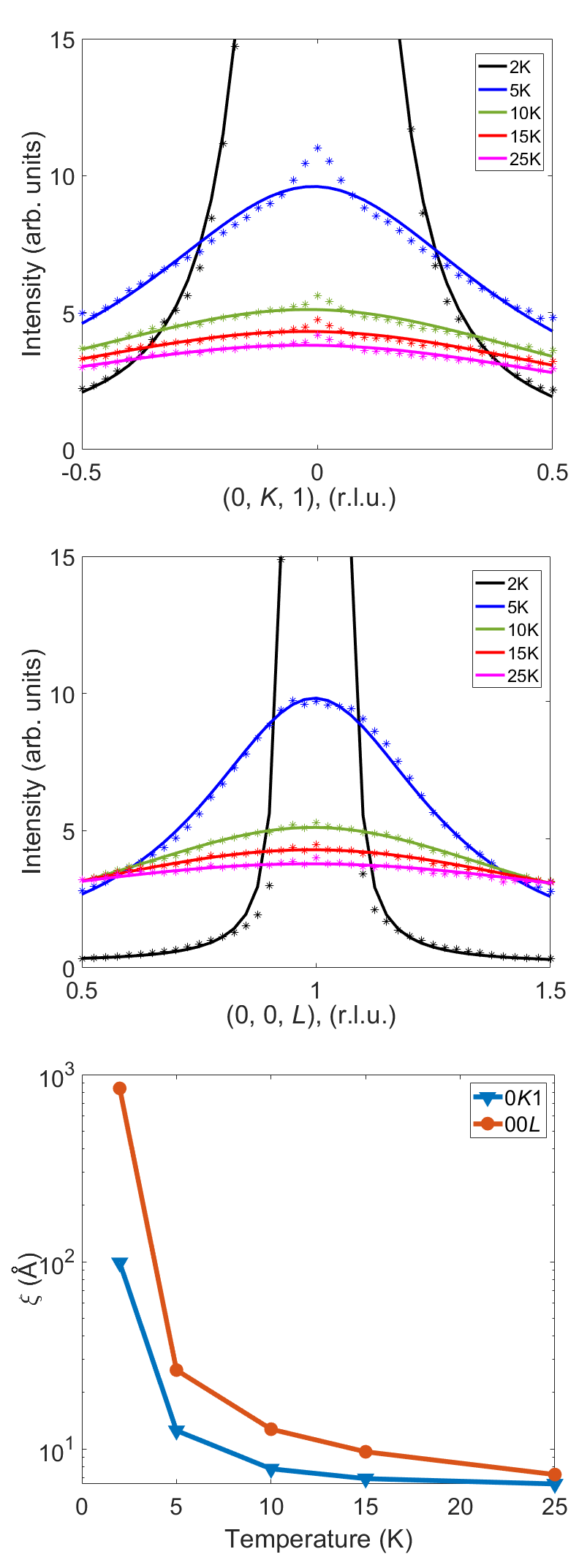} b\\
	\end{minipage}
	
	\caption{ (a) Experimental data integrated over the H directions of the reciprocal space in the limit ±0.1 r.l.u. The energy axis was integrated within ±0.1 meV. From up to down located slices, maken from different temperature. (b) Experimental data integrated over the directions of the reciprocal space in the limit ±0.15 r.l.u. The energy axis was integrated within ±0.1 meV. Result of the approximation of the experimental data by the convolution of the Lorentzian and Gaussian functions. The resolution was estimated from the structural peak width using two characteristic directions of the reciprocal space. On the last figure temperature dependence of the correlation length for both directions of the reciprocal space.}\label{Corlength} 
	
\end{figure}

Having considered the energy range below 3.3 meV, we investigated the magnetic excitations of the Tb subsystem and determined the correlation lengths in two directions. The intensity cuts of the detected elastic peaks are shown in Fig. ~\ref{Corlength}(a). The examination of the low-energy spectra at the diffuse scattering allowed us to describe the magnetic structure factor. Here, as in ~\cite{39, 40}, we used the Lorentzian function

\begin{equation}
	S(Q)\propto\frac{sinh(c/\xi_l)}{cosh(c/\xi_l)-cos[\pi(l-1)]}\frac{sinh(c/\xi_k)}{cosh(c/\xi_k)-cos[\pi k]}
\end{equation}
where $\xi_l$ and $\xi_k$ are the correlation lengths along the c and b axes, respectively.

Making the energy cuts in the K and L directions of the reciprocal space, we obtained the data shown by dots in Figs. ~\ref{Corlength}b. The approximation of the experimental data using the structural peak width and the Lorentzian function yielded the correlation lengths for each direction and temperature. The temperature dependence of the correlation length for the two investigated directions is shown in Fig. ~\ref{Corlength}(b). It can be clearly seen that, after the spin-reorientation transition, the correlation length sharply increases, which is consistent with the fact of ordering of both subsystems below this transition.

\section{DISCUSSION AND CONCLUSION}

The inelastic neutron scattering study of the single-crystal TbFeO$_3$ sample showed the antiferromagnetic ordering of the Fe$^{3+}$ subsystem. Using the spectra obtained, the Fe‒Fe exchange interaction values were calculated within the linear spin wave theory. The spectra are similar to the spectra obtained for iron in the TmFeO${_3}$ compound [24], which manifests itself in the energy levels of the magnon dispersion, in the exchange interaction values, and, importantly, in the anisotropy of the exchange values determined for these compounds. The antiferromagnetic gap observed in the investigated compound at 4.5 meV is governed by the uniaxial anisotropy, which was also established by us in the framework of the linear spin wave theory. The gap depends on the type of a rare-earth ion contained in the compound.

The examination of the spectra in the range of up to 50 meV elucidated the behavior of the Tb3+ subsystem, specifically, the position of the first three CEF levels, which are located at energies of 17, 25, and 35 meV for the compound under study. We failed to establish any dispersion behavior of these levels from the experimental data; it was found that the first level has the highest population at the investigated temperature and the intensity of this level competes with that of Fe$^{3+}$ magnons.

The magnetic subsystem of the rare-earth Tb$^{3+}$ ion was described using the point charge model. In this model, the magnetization of the terbium subsystem and the magnon dispersion were calculated and plotted. The comparison of the calculation with the experiment demonstrated their good agreement. The correlation lengths at different temperatures were determined, which made it possible to establish the temperature above which the terbium subsystem acquires a long-range magnetic order.

\section{ACKNOWLEDGMENTS}
We acknowledge A. Podlesnyak for help with INS experiments and S. E. Nikitin for stimulating discussion.
This study was supported by the Russian Foundation for Basic Research, project no. 20-32-90142.
This research used resources at Spallation Neutron Source, a DOE Office of Science User Facility operated by the Oak Ridge National Laboratory. 

\section{APPENDIX:RESULTS OF THE POINT-CHARGE MODEL CALCULATIONS}

The set of CEF parameters and the energy levels calculated from the PC model are given in Tables A1 and A2, respectively.

\tablename{ A1. Calculation of Stevens parameters B$^m_l$ and energy levels with corresponding magnetic moments.} \label{tab:blm}

\small

\begin{tabular}{llcllc}
	
	\hline\hline
	
	$B_l^m$&   &   & $B_l^m$ &   &   \\
	
	\hline
	
	$B_2^0$ & = & $3.91\times10^{-1}$ & $B_6^0$ & = & $0.1\times10^{-5}$\\
	
	$B_2^2$ & = & $0.39\times10^{-1}$ & $B_6^2$ & = & $-0.3\times10^{-5}$\\
	
	$B_2^{-2}$ & = & $-3.03\times10^{-1}$ & $B_6^-2$ & = & $-0.3\times10^{-5}$ \\
	
	$B_4^0$ & = & $-0.35\times10^{-3}$ & $B_6^4$ & = & $1.8\times10^{-5}$\\
	
	$B_4^2$ & = & $-1.34\times10^{-3}$ & $B_6^-4$ & = &  $0.1\times10^{-5}$\\
	
	$B_4^{-2}$ & = & $-3.37\times10^{-3}$ & $B_6^6$ & = & $-0.3\times10^{-5}$ \\
	
	$B_4^4$ & = & $0.45\times10^{-3}$ & $B_6^-6$ & = & $-0.0\times10^{-5}$\\
	
	$B_4^{-4}$ & = & $-8.1\times10^{-3}$&  &  &  \\
	\hline
	
\end{tabular}

\

\

\tablename{ A2. The energy levels and out of ground state transition probabilities.}

\small

\begin{tabular}{lcccc}

	\hline\hline
	
	& $E$ (meV) & $\langle n|J_x|m\rangle^2$ & $\langle n|J_y|m\rangle^2$ & $\langle n|J_z|m\rangle^2$ \\
	
	\hline
	
	$|E_0\rangle\to|E_1\rangle$ & 0.0169 & 0.0 & 0.0 & 35.0109 \\
	
	$|E_0\rangle\to|E_2\rangle$ & 13.9161 & 0.0 & 3.0388 & 0.1414 \\
	
	$|E_0\rangle\to|E_3\rangle$ & 14.1608 & 0.8099 & 0.0 & 0.0 \\
	
	$|E_0\rangle\to|E_4\rangle$ & 24.7651 & 1.1200 & 0.0 & 0.0 \\
	
	$|E_0\rangle\to|E_5\rangle$ & 25.7851 & 0.0 & 1.4517 & 0.0658 \\
	
	$|E_0\rangle\to|E_6\rangle$ & 34.3537 & 0.0 & 0.0881 & 0.0154 \\
	
	$|E_0\rangle\to|E_7\rangle$ & 34.7824 & 0.325 & 0.0 & 0.0 \\
	
	$|E_0\rangle\to|E_8\rangle$ & 39.0313 & 0.0 & 0.0019 & 0.0052 \\
		
	$|E_0\rangle\to|E_9\rangle$ & 43.7716 & 0.0011 & 0.0 & 0.0 \\
		
	$|E_0\rangle\to|E_{10}\rangle$ & 44.4169 & 0.0319 & 0.0 & 0.0 \\
		
	$|E_0\rangle\to|E_{11}\rangle$ & 53.4470 & 0.0 & 0.0013 & 0.0007 \\
		
	$|E_0\rangle\to|E_{12}\rangle$ & 53.4992 & 0.0 & 0.0007 & 0.0003 \\
	\hline
	
\end{tabular}


\vspace{10pt}


\begin{thebibliography}{99}

\bibitem{1} Salamon, Myron B., and Marcelo Jaime, The physics of manganites: Structure and transport, Rev. Mod. Phys. \textbf{73}, 583, (2001)

\bibitem{2} Tokura, Yoshinori, and Yasuhide Tomioka, Colossal magnetoresistive manganites, J. Magn. Magn. Mater. \textbf{200}, 1, (1999).

\bibitem{3} Kimura, Hiroyuki, et al, Spiral Spin Structure in the Commensurate Magnetic Phase of Multiferroic RMn$_2$O$_5$, J. Phys. Soc. Jpn. \textbf{76}, 074706,  (2007).

\bibitem{4} Schieber, M., et al., Magnetocrystalline anisotropy of rare‐earth manganites, J.Appl. Phys. \textbf{44}, 1864 (1973).

\bibitem{5} Ritter, C., et al., Magnetic structure in iron borates RFe$_3$(BO$_3$)$_4$ (R=Y, Ho): a neutron diffraction and magnetization study, J. Phys.: Condens. Matter, \textbf{20}, 365209, (2008).

\bibitem{6} Pankrats, A. I., et al., Transformation from an easy-plane to an easy-axis antiferromagnetic structure in the mixed rare-earth ferroborates Pr$_x$Y$_{1-x}$Fe$_3$(BO$_3$)$_4$: magnetic properties and crystal field calculations, J. Phys.: Condens. Matter, \textbf{28}, 396001, (2016).

\bibitem{7} Krylov, A. S., et al., Manifestation of magnetoelastic interactions in Raman spectra of Ho$_x$Nd$_{1-x}$Fe$_3$(BO$_3$)$_4$ crystals, J. Adv. Dielectr., \textbf{08}, 1850011, (2018).

\bibitem{8} Pankrats, A., et al., Low-temperature magnetic phase diagram of HoFe$_3$(BO$_3$)$_4$ holmium ferroborate: a magnetic and heat capacity study, J. Phys.: Condens. Matter  \textbf{21}, 436001, (2009).

\bibitem{9}	R. L. White, Work on the Magnetic and Spectroscopic Properties of the Rare-Earth Orthoferrites, J. Appl. Phys. \textbf{40}, 1061, (1969). 

\bibitem{10} Sosnowska, I., and A. K. Zvezdin, Origin of the long period magnetic ordering in BiFeO$_3$, J. Magn. Magn. Mater. \textbf{140}, 167, (1995).

\bibitem{11} Ruette, Benjamin, et al., Magnetic-field-induced phase transition in BiFeO$_3$ observed by high-field electron spin resonance: Cycloidal to homogeneous spin order, Phys. Rev. B \textbf{69}, 064114, (2004).

\bibitem{12} K. P. Belov, A. K. Zvezdin, and A. A. Mukhin, Magnetic phase transitions in terbium orthoferrite, J. Exp. Theor. Phys. \textbf{49}, 557, (1979).

\bibitem{13} Tokunaga, Yusuke, et al., Composite domain walls in a multiferroic perovskite ferrite, , Nature Mater. \textbf{8}, 558 (2009).

\bibitem{14} Kimel, A. V., et al., Ultrafast non-thermal control of magnetization by instantaneous photomagnetic pulses, Nature \textbf{435}, 655, (2005).

\bibitem{15} De Jong, J. A., et al., Laser-induced ultrafast spin dynamics in ErFeO$_3$, Phys. Rev. B, \textbf{84}, 104421, (2011).

\bibitem{16} Jiang, Junjie, et al., Dynamical spin reorientation transition in NdFeO$_3$ single crystal observed with polarized terahertz time domain spectroscopy, Appl. Phys. Lett. \textbf{103}, 062403, (2013).

\bibitem{17} S. Artyukhin, M. Mostovoy, N. P. Jensen, D. Le, K. Prokes, V. G. de Paula, H. N. Bordallo, A. Maljuk, S. Landsgesell, H. Ryll, B. Klemke, S. Paeckel, K. Kiefer, K. Lefmann, L. T. Kuhn, and D. N. Argyriou, Solitonic lattice and Yukawa forces in the rare-earth orthoferrite TbFeO$_3$, Nat. Mater. \textbf{11}, 694 (2012)

\bibitem{18} Saito, Kazuya, et al., High-precision detection of the heat-capacity anomaly due to spin reorientation in TmFeO3 and HoFeO3, Solid State Commun. \textbf{120}, 129, (2001).

\bibitem{19} O Nikolov et al., A Mossbauer study of temperature-driven spin-reorientation transitions in TbFeO$_3$, J. Phys.: Condens. Matter, \textbf{6}, 3793, (1994).

\bibitem{20} Dzialoshinskii, I. E., Thermodynamic theory of weak ferromagnetism in antiferromagnetic substances, Soviet Physics Jetp-Ussr, \textbf{5}, 1259, (1957).

\bibitem{21} A Podlesnyak et al., Low-energy spin dynamics in rare-earth perovskite oxides, J. Phys.: Condens. Matter, \textbf{33}, 403001, (2021).

\bibitem{22} S. E. Nikitin et al., Decoupled spin dynamics in the rare-earth orthoferrite YbFeO$_3$: Evolution of magnetic excitations through the spin-reorientation transition, Phys. Rev. B \textbf{98}, 064424, (2018).

\bibitem{23} L.S. Wu et al., Tomonaga-Luttinger liquid behavior and spinon confinement in YbAlO$_3$, Nat. Commun. \textbf{10}, 698, (2019).

\bibitem{24} S.E Nikitin et al., Multiple fermion scattering in the weakly coupled spin-chain compound YbAlO$_3$, Nat. Commun. \textbf{12}, 1, (2021).

\bibitem{25} S. A. Skorobogatov et al., Low-temperature spin dynamics in the TmFeO$_3$ orthoferrite with a non-Kramers ion, Phys. Rev. B \textbf{101}, 014432, (2020).

\bibitem{26} Y. Cao et al., Magnetic phase transition and giant anisotropic magnetic entropy change in TbFeO$_3$ single crystal, J. Appl. Phys. \textbf{119}, 063904, (2016).

\bibitem{27} D. L. Abernathyet et al., Design and operation of the wide angular-range chopper spectrometer ARCS at the Spallation Neutron Source, Rev. Sci. Instrum. \textbf{83}, 015114, (2012).

\bibitem{28} G. Ehlers et al., The new cold neutron chopper spectrometer at the spallation neutron source: design and performance, Rev. Sci. Instrum. \textbf{82}, 085108, (2011).

\bibitem{29} G. Ehlers, A. Podlesnyak, and A. I. Kolesnikov, The cold neutron chopper spectrometer at the Spallation Neutron Source - A review of the first 8 years of operation, Rev. Sci. Instrum. \textbf{87}, 093902, (2016).

\bibitem{30} R. Azuah et al., DAVE: a comprehensive software suite for the reduction, visualization, and analysis of low energy neutron spectroscopic data, J. Res. Natl. Inst. Stan. Technol. \textbf{114}, 341, (2009).

\bibitem{31} R. A. Ewings, et al., HORACE: software for the analysis of data from single crystal spectroscopy experiments at time-of-light neutron instruments, Nucl. Instrum. Methods Phys. Res. Sect. \textbf{A 834}, 132, (2016).

\bibitem{32} O. Arnold et al., Mantid – Data analysis and visualization package for neutron scattering and $\mu$SR experiments, Nucl. Instrum. Methods Phys. Res. Sect. \textbf{A 764}, 156, (2014).

\bibitem{33} S. Toth and B. Lake, Linear spin wave theory for single-Q incommensurate magnetic structures, J. Phys.: Condens. Matter \textbf{27}, 166002, (2015).

\bibitem{34} M Rotter, Using mcphase to calculate magnetic phase diagrams of rare earth compounds, J. Magn. Magn. Mater. \textbf{272}, E481, (2004).

\bibitem{35} S. Hahn et al., Inelastic neutron scattering studies of YFeO$_3$, Phys. Rev. B \textbf{89}, 014420, (2014).

\bibitem{36} S. M. Shapiro, J. D. Axe, and J. P. Remeika, Neutronscattering studies of spin waves in rare-earth orthoferrites, Phys. Rev. B \textbf{10}, 2014, (1974).

\bibitem{37} To stabilize the correct ground state with a small spin canting ~\cite{41} we have to introduce two Dzyaloshinskii- Moriya interaction constants to Eq. ~\ref{Hamiltonian}. However, both are small and have a negligible effect of the spectra.

\bibitem{38} M. T. Hutchings, Point-charge calculations of energy levels of magnetic ions in crystalline electric fields, 1964, Solid State Phys., \textbf{16}, 227–273.

\bibitem{39} I. Zaliznyak et al., Spin-liquid polymorphism in a correlated electron system on the threshold of superconductivity, Proc. Natl. Acad. Sci. (USA) \textbf{112}, 10316, (2015).

\bibitem{40} L. S. Wu et al., Magnetic ground state of the Ising-like antiferromagnet DyScO$_3$, Phys. Rev. B \textbf{96}, 144407, (2017).

\bibitem{41} R. L. White, Work on the Magnetic and Spectroscopic
Properties of the Rare-Earth Orthoferrites, J. Appl. Phys.
\textbf{40}, 1061 (1969).


\end{thebibliography}
\end{document}